\documentclass{jnmp}
\relax
\usepackage{amsmath}

\JNMPnumberwithin{equation}{section}

\theoremstyle{definition}

\begin{document}
\renewcommand{\evenhead}{B. A. Kupershmidt}
\renewcommand{\oddhead}{Equations Of Long Waves With A Free Surface IV.  The Case of Constant Shear }
\thispagestyle{empty}

\Name{EQUATIONS OF LONG WAVES WITH A FREE SURFACE IV.  THE CASE OF CONSTANT SHEAR}
\label{firstpage}

\Author{Boris A. Kupershmidt~$\dag$}
\Address{$^\dag$ The University of Tennessee Space Institute \\
~~Tullahoma, TN  37388, USA \\
~~E-mail:  bkupersh@utsi.edu\\[10pt]}


\begin{abstract}
\noindent
A large class of two-dimensional free-surface hydrodynamical systems is determined that can be self-consistently 
reduced by the condition that the velocity profile has a constant shear.  The reduced systems turn out to be 
Hamiltonian, and so does the reduction process itself. All reducible systems, Hamiltonian or not, are determined and 
shown to form a Lie algebra.  All this is then generalized to the multilayer/multi-species representations.
\end{abstract}


\section{Introduction}  

The classical one-dimensional long-wave system
\begin{subequations} \label{eqA}
\begin{gather}
h_t = (hu)_x, \\
u_t = uu_x + gh_x,
\end{gather}
\end{subequations}
has a venerable history.  Here \(h=h(x,t)\) is the height of a free surface over the bottom \( \{y=0\}; \ u = 
u (x, t)\) is the horizontal component of velocity; $t$ is the time coordinate (opposite in sign to the 
physical time); $- \infty < x < \infty$; subscripts $t$ and $x$ denote partial derivatives; $g$ is the gravitational 
acceleration.

The system (1.1) is Hamiltonian and integrable:  it can be put into the form [5,6]
\begin{equation}
\binom{h}{u}_t \ = \ \binom{0 \ \partial} {\partial \ 0} \ 
\binom{\delta H/\delta h} {\delta H/\delta u}, 
\end{equation}
where
\begin{gather}
\partial = \partial/\partial x, \\
H = \frac {1} {2} (hu^2 + gh^2),  
\end{gather}
and there exists an infinite number of conserved densities for that system.

In 1973 Benney [1] derived the following {\it{two-dimensional}} generalization of the system (1.1):

\begin{subequations} 
\begin{gather}
h_t = (\int\limits^{h}\limits_{0} udy)_x,  \\
u_t = uu_x + gh_x - u_y \int\limits^{y}\limits_{0} u_x dy, 
\end{gather}
\end{subequations}
where now \(u = u (x, y, t)\) depends also upon the second space coordinate, \(y: 0 \leq y \leq h\).  Benney had found 
two remarkable properties of the {\it{two-dimensional}} system (1.5).  First, if one introduces the {\it{moments}} 
of the velocity \(u (x, y, t)\):
\begin{equation}
A_n = A_n (x, t) = \ \int^n_0 u^n (x, y, t) dy, \ \ \ n \in {\bf{Z}}_{\geq 0}, 
\end{equation}
then the system (1.5) implies the {\it{autonomous}} evolution system
\begin{equation}
A_{n,t} = A_{n+1,x} + gn A_{n-1} A_{0,x}, \ \ \ n \in {\bf{Z}}_{\geq 0}. 
\end{equation}

Second, the moments system (1.7) has an {\it{infinite}} number of polynomial conserved densities \(H_n \in A_n + 
{\bf{Q}} [g; A_0, ..., A_{n-2}] \) : 

\begin{equation}
H_0 = A_0, \ H_1 = A_1, \ \ H_2 = A_2 + g A_{0}^{2}, ... 
\end{equation}
\\
\noindent
Subsequently, Manin and myself showed [6,7] that:  
\\

(A) \ The moment system (1.7) is itself Hamiltonian: it can be written in the form

\begin{gather}
A_{n,t} = \sum_{m \geq 0} B_{nm} (H_m), \ \ \ H_m = \frac {\delta H} {\delta A_m},  \\
B_{nm} = n A_{n+m-1} \partial + \partial m A_{n+m-1}, 
\end{gather}
with \(H = \frac {1} {2} H_2 = \frac {1} {2} (A_2 + g A^2_0)\), and with the matrix (1.10) being Hamiltonian; 

(B) \ The general system (1.9) is implied by the following two-dimensional free-surface system:

\begin{subequations} 
\begin{gather}
h_t = \partial (m A_{m-1} H_m) \\
u_t = \partial (u^{m} H_m) - u_y \int\limits^{y}\limits_{0} dy (mu^{m-1} H_m)_x.  
\end{gather}
\end{subequations}

We sum on repeated non-fixed indices unless directed otherwise; 

(C) \ The Hamiltonians $H_{k}$'s (1.8) found by Benney are in involution with respect to the Hamiltonian structure 
(1.10).  Therefore, the corresponding higher flows commute;

(D) \ When $u$ is $y$-{\it{independent}}, 

\begin{equation}
u_y = 0, 
\end{equation}
so that we are back to the classical one-dimensional case, the map (1.6) becomes

\begin{equation}
A_n = h u^n, \ \ \ n \in {\bf{Z}}_{\geq 0}, 
\end{equation}
and this map is {\it{Hamiltonian}} between the Hamiltonian structures (1.2) and (1.10).

The purpose of this note is to show that there exists an interesting reduced family of the full two-dimensional 
system (1.11) generalizing the purely one-dimensional reduction \(u_y = 0\) (1.12), namely

\begin{equation}
u_y = s = \ const, 
\end{equation}
\begin{equation}
u(x, y, t) = v(x, t) + sy, \ \ \ \ s = \ const. 
\end{equation}
Thus, we consider the case when the shear {\it{is}} present but is constant.

We shall verify that:  the constraint \(\{u_y = s\}\) (1.14) is compatible with the flow (1.11) for {\it{any}} 
Hamiltonian $H$; that on this constrained submanifold \(\{u_y = s\}\) (1.15), the system (1.11) turns into a 
Hamiltonian system of the form 
\begin{equation}
\binom{h_t}{v_t} = \binom{0 \ \ \ \ \ \partial}{\partial \ \ -s \partial}
\binom{\delta H/\delta h}{\delta H/\delta v}; 
\end{equation}
and that the corresponding reduction map 

\begin{equation}
A_n = \int\limits^{h}\limits_{0} (v + sy)^n dy, \ \ \ \ n \in {\bf{Z}}_{\geq 0}, 
\end{equation}
is Hamiltonian between the Hamiltonian structures (1.16) and (1.10).

We then determine when similar constant-shear reductions exist for other free-surface hydrodynamical systems.

At the moment, let us record that the original two-dimensional Benney system (1.5) reduces on the submanifold \(
\{u_y = s, \ u = v + sy\},\) to the system

\begin{equation}
\left(
\begin{array}{c}
h_t \\ 
\\ 
\\
v_t 
\end{array} \right)
= 
\left(
\begin{array}{c}
(vh + s \frac {h^2} {2})_x \\ 
\\
\\
vv_x \ + \ gh_x 
\end{array} \right)
=
\left(
\begin{array}{cc}
0 & \partial \\
\\
\\
\partial & - s \partial 
\end{array} \right)
\left(
\begin{array}{c}
{\displaystyle{\frac{(v+sh)^2} {2}}} + gh \\
\\ 
\\
vh + s {\displaystyle{\frac {h^2}{2}}} \\
\end{array}  \right),  \tag{1.18}
\end{equation}

\begin{subequations} 
\begin{gather*}
H = \frac {1} {2} (A^*_2 + g A^{*2}_0) = \frac {1}{2} \bigg( \frac {(v+ sh)^3 - v^3} {3s} + gh^2 \bigg),  
\tag{1.19$a$} \\ 
\frac {\delta H} {\delta h} = \frac {(v+sh)^2} {2} + gh, \ \ \frac {\delta H} {\delta v} = vh + s \frac 
{h^2}{2}.\tag{1.19$b$}
\end{gather*}
\end{subequations}

\section{Constant-{\it{\bf{Shear Flows}}}}  

Denote by \( ( \cdot )^* \) the reduction of the object \( ( \cdot ) \) on the submanifold

\begin{equation}
u_y = s, \ u = v+sy. 
\end{equation}
Thus,

\begin{gather*}
A^*_m = \int\limits^{h}\limits_{0} (v + sy)^m dy = \frac {(v+sh)^{m+1} - v^{m+1}} {s(m+1)}, \ \ \ s \neq 0, \tag{2.2} 
\\
A^*_{m,v} = \frac {\partial A^*_m} {\partial v} = mA^*_{m-1} = \frac {(v+sh)^m - v^m} {s}, \tag{2.3} \\
A^*_{m,h} = \frac {\partial A^*_m} {\partial h} = (v + sh)^m, \tag{2.4} \\
H^*_h = \frac {\delta (H^*)} {\delta h} = \bigg( \frac {\delta H} {\delta A_m} \bigg)^* \frac {\partial A^*_m} 
{\partial h} = (v + sh)^m H^*_m \tag{2.5} \\
H^*_v = H^*_m \frac{\partial A^*_m} {\partial v} = (m A_{m-1} H_m)^* = \tag{2.6$a$} \\
= \frac{(v + s h)^m - v^m} {s} H^*_m . \tag{2.6$b$} \\
\end{gather*}

Now, denote temporarily

\begin{equation}
F = u^m H_m, \tag{2.7}
\end{equation}
so that
\begin{equation}
F_u = mu^{m-1} H_m. \tag{2.8}
\end{equation}
Differentiating equation (1.11$b$) with respect to $y$, we find:
\begin{gather}
u_{y,t} = \partial (F_u u_y) - u_{yy} \int\limits^{y}\limits_{0} dy (F_u)_x - u_y (F_u)_x = \notag \\
= F_u u_{yx} - u_{yy} \int\limits^{y}\limits_{0} dy (F_u)_x. \tag{2.9}
\end{gather}
When \(u_y = s = const, u_{yt} = u_{yx} = u_{yy} = 0\).  Thus, the flow (1.11) properly restricts on the 
constraint \( \{u_y = s, \ u= v+sy \}\).  Evaluating equation (1.11$b$) at $y=0$, we obtain:
\begin{equation}
v_t = \partial (v^m H_m^*). \tag{2.10}
\end{equation}
Equation (1.11$a$) becomes:
\begin{equation}
h_t = \partial (m A_{m-1} H_m)^* \ {\rm{[by \ (2.6a)]}} \ = \partial \bigg( \frac {\delta H^*} {\delta v} \bigg), 
\tag{2.11}
\end{equation}
which proves the first half of formula (1.16).  To prove the second half of that formula, we need to check, 
in view of the relation (2.10), that
\begin{equation}
\partial (v^m H^*_m) \ \shortstack{?\\=} \ \partial (H_h^*) - s \partial (H^*_v) \tag{2.12}
\end{equation}
or
\begin{equation}
v^m H^*_m \ \shortstack{?\\=} \ H^*_h - s H^*_v. \tag{2.13}
\end{equation}
By formulae (2.5,6$b$), we have to verify that

\begin{equation}
v^m H^*_m \ \shortstack{?\\=} \ (v + sh)^m H^*_m - \bigg( (v+sh)^m - v^m\bigg) H^*_m, \tag{2.14}
\end{equation}
which is obviously true. 

\section{The Reduction Map Is Hamiltonian}

We need to verify that the map (2.2), 

\begin{equation}
A^*_m = \frac{(v+sh)^{m+1} - v^{m+1}} {s(m+1)}, \ \ \ m \in {\bf{Z}}_{\geq 0}, 
\end{equation}
is Hamiltonian between the Hamiltonian matrices
\begin{equation}
b = \binom{0 \ \ \ \ \ \ \partial}{\partial \ \ -s \partial} 
\end{equation}
and
\begin{equation}
B_{nm} = n A_{n+m-1} \partial + \partial m A_{n+m-1}. 
\end{equation}
This is equivalent to the equality
\begin{equation}
B^* \ \shortstack{?\\=} \ J b J^t, 
\end{equation}
where $J$ is the Fr\'{e}chet Jacobian of the map (3.1):
\begin{equation}
J_{n,h} = \frac {\partial A_n^*}{\partial h} = A^*_{n,h}, \ J_{n,v} = \frac {\partial A^*_n} {\partial v} = 
A^*_{n,v}. 
\end{equation}
In components, the equality (3.4) becomes:
\begin{gather}
n A^*_{n+m-1} \partial + \partial m A_{n+m-1}^* \ \shortstack{?\\=} \notag \\
= A^*_{n,v} \partial A^*_{m,h} + (A^*_{n,h} - s A^*_{n,v}) \partial A^*_{m,v} , \ \ \ n, m \in {\bf{Z}}_{\geq 0}. 
\end{gather}
This identity in turn, splits into the pair:
\begin{gather}
(n+m) A_{n+m-1}^* \ \shortstack{?\\=} \ A^*_{n,v} A^*_{m,h} + (A^*_{n,h} - s A^*_{n,v}) A^*_{m,v},  \\
m A^*_{n+m-1,x} \ \shortstack{?\\=} \ A^*_{n,v} (A^*_{m,h})_x + (A^*_{n,h} - s A^*_{n,v}) (A^*_{m,v})_x. 
\end{gather}

We start with the identify (3.7).  From formulae (2.3,4) we have:

\begin{equation}
A^*_{n,h} - s A^*_{n,v} = v^n. 
\end{equation}
Denote
\begin{equation}
\sigma = v + sh. 
\end{equation}
Formula (3.7) becomes in view of formula (2.2):
\begin{equation}
(n+m) \frac {\sigma^{n + m} - v^{n+m}} {s(n+m)} \ \shortstack{?\\=} \ \frac {\sigma^n - v^n} {s} \sigma^m + v^n 
\frac {\sigma^m - v^m} {s},  
\end{equation}
which is obviously true.

Next, formula (3.8) becomes:
\begin{equation}
m \bigg ( \frac {\sigma^{n+m} - v^{n+m}} {s(n+m)}\bigg)_x \ \shortstack{?\\=} \ \frac {\sigma^n - v^n} {s}
(\sigma^m)_x + v^n \bigg( \frac {\sigma^m - v^m} {s} \bigg)_x. 
\end{equation}
Since
\begin{equation}
\frac {\partial \sigma} {\partial v} = 1, 
\end{equation}
the identity (3.12) splits into the pair:  
\begin{gather}
m \sigma^{n+m-1} \sigma_h h_x \ \shortstack{?\\=} \ (\sigma^n - v^n) m \sigma^{m-1} \sigma_h h_x + v^n m \sigma^{m-1}
\sigma_h h_x,  \\
m (\sigma^{n+m-1} - v^{n+m-1}) v_x \ \shortstack{?\\=} \ (\sigma^n - v^n) m \sigma^{m-1} v_x + v^n m (\sigma^{m-1} 
- v^{m-1}) v_x, 
\end{gather}
and each one of these identities is obviously true. 
\section{Other Two-Dimensional Systems}

Free-surface systems, such as (1.11), are naturally attached to local Lie algebras, in particular to Poisson 
manifolds [4].  In the two-dimensional case, the general form of systems liftable into the space of moments has 
the form [4]

\begin{subequations} 
\begin{gather}
h_t = Q_m A_m + P_m A_{m,x}  \\
u_t = P_m u^m u_x + \bar P_m u^m - u_y \int\limits^{y}\limits_{0} dy (P_m (u^m)_x + Q_m u^m), 
\end{gather}
\end{subequations}
where $P_m, \bar P_m, Q_m$ are arbitrary functions of $x$ and the $A_n$'s; the resulting evolution for the 
moments is:

\begin{equation}
A_{n,t} = n A_{n+m-1} \bar P_m + A_{n+m} Q_m + A_{n+m,x} P_m. 
\end{equation}

The systems (1.11) we have looked at in the previous Sections are of the above form, with

\begin{equation}
\bar P_m = H_{m,x} , \ Q_m = (m+1) H_{m+1,x}, \ \ P_m = (m+1) H_{m+1}. 
\end{equation}

Let us determine when the system (4.1) can be self-consistently constrained onto the submanifold \( \{u_y = s\}\). 
Differentiating formula (4.1$b$) with respect to $y$, we get:

\begin{subequations} 
\begin{gather}
u_{y,t} = P_m (mu^{m-1} u_y u_x + u^m u_{xy} ) + \bar P_m mu^{m-1}u_y - \notag \\
- u_{yy} \int\limits^{y}\limits_{0} dy (P_m (u^m)_x + Q_m u^m) - u_y (P_m mu^{m-1} u_x + Q_m u^m) = \notag \\
= P_m u^m u_{xy} - u_{yy} \int\limits^{y}\limits_{0} dy (P_m (u^m)_x + Q_m u^m) +  \\
+ u_y u^m ((m+1) \bar P_{m+1} - Q_m). 
\end{gather}
\end{subequations}
Hence, the system (4.1) is constrainable iff
\begin{equation}
Q_m = (m+1) \bar P_{m+1}, \ \ m \in {\bf{Z}}_{\geq 0}. 
\end{equation}
\noindent
The resulting $h,v$-system can be read off formulae (4.1) for $y=0$:
\begin{gather}
h_t = (m+1) \bar P^*_{m+1} A^*_m + P^*_m A^*_{m,x}, \tag{4.6$a$}\\
v_t = P_m^* v^m v_x + \bar P_m^* v^m. \tag{4.6$b$} \
\end{gather}

Formula (4.3) shows that the relations (4.5) are satisfied for our original system (1.11).
\

The system (4.1) is of a {\it{general character}}.  Among {\it{Hamiltonian}} systems of this type, there exists 
a two-parameter family [2, formula (2.99$^\prime$)] given by the Hamiltonian matrix
\begin{equation}
B^{\alpha, \beta}_{nm} = (\alpha n + \beta) A_{n+m} \partial + \partial (\alpha m+ \beta) A_{n+m}, \tag{4.7}
\end{equation}
where $\alpha$ and $\beta$ are arbitrary constants.  For this case, we have:

\begin{gather}
A_{n,t} = B^{\alpha, \beta}_{nm} (H_m) = (\alpha n + \beta) A_{n+m} H_{m,x} + (\alpha m + \beta ) (A_{n+m} H_m)_x = 
\notag \\
= n A_{n+m} \alpha H_{m,x} + A_{n+m} (\alpha m + 2 \beta) H_{m, x} + A_{n+m,x} (\alpha m + \beta) H_m . \tag{4.8}\
\end{gather}
From formula (4.2) we see that
\begin{equation}
\bar P_0 = 0; \ \bar P_{m+1} = \alpha H_{m,x}; \ Q_m = (\alpha m + 2 \beta) H_{m,x} ; \ P_m = (\alpha m+\beta ) H_m 
.\tag{4.9} 
\end{equation}
Therefore, the constrainability criterion (4.5) is satisfied provided
\begin{equation}
(\alpha m + 2 \beta) H_{m,x} = (m + 1) \alpha H_{m,x}, \ \ \ m \in {\bf{Z}}_{\geq 0}, \tag{4.10$a$} 
\end{equation}
or
\begin{equation}
(\alpha m + 2 \beta) = (m + 1) \alpha, \ \ \ m \in {\bf{Z}}_{\geq 0}, \tag{4.10$b$}
\end{equation}
and this happens iff
\begin{equation}
\alpha = 2 \beta. \tag{4.11}
\end{equation}

This is a very puzzling result.  To see why, notice that the Hamiltonian matrix $B^{\alpha, \beta}$ (4.7) is 
{\it{linear}} in the field variables (the $A_n$'s). Hence, it corresponds to a Lie algebra.  An easy calculation 
shows that this Lie algebra has the commutator

\begin{equation}
[{\boldsymbol{X}}, {\boldsymbol{Y}}]_k = \sum_{n + m=k} ((\alpha n + \beta) X_n Y_{m,x} - (\alpha m + \beta) Y_m 
X_{n,x}). \tag{4.12}
\end{equation}

Setting

\begin{equation}
f (x, p) = \sum_{n \geq 0} X_n p^n , \ \ \ g (x, p) = \sum_{m \geq 0} Y_m p^m, \tag{4.13} 
\end{equation}
we can convert the commutator (4.12) into the following Poisson bracket on \({\bf{R}}^2\):

\begin{equation}
\{ f, g \} = \beta (f g_x - f_x g) + \alpha p (f_p g_x - f_x g_p). \tag{4.14}
\end{equation}
There is {\it{nothing}} in this Poisson bracket to indicate that the ratio

\begin{equation}
\alpha: \beta = 2:1 \tag{4.15}
\end{equation}
is distinguished from all the other ratios.

Let us now consider what happens with system (4.8) for the case
\begin{equation}
\alpha = 2, \ \beta = 1, \tag{4.16}
\end{equation}
when this system is restricted onto the submanifolds $\{ u_y = s \}$.   By formula (4.9), the full system (4.1) 
has the form:
\begin{gather}
h_t = A_m H_{m,x} + (2 m+1) (A_m H_m)_x, \tag{4.17$a$} \\
u_t = (2 m+1) u^m u_x H_m + 2u^{m+1} H_{m,x} - u_y \int\limits^{y}\limits_{0} dy ((2 m+1) (u^m)_x H_m + \notag \\
+ 2 (m+1) u^m H_{m,x}). \tag{4.17$b$} \
\end{gather}

Hence, the restricted system becomes:

\begin{gather}
h_t = A^*_m H^*_{m,x} + (2m+1) (A^*_m H^*_m)_x,  \tag{4.18$a$} \\
v_t = (2m+1) v^m v_x H^*_m + 2v^{m+1} H^*_{m,x}. \tag{4.18$b$} \
\end{gather}

{\bf{Proposition 4.19.}} \ (i) The system (4.18) can be put into the following form:

\begin{equation}
\left(
\begin{array}{c}
h \\
\\
\\
v \\
\end{array} \right)_t
=
\left(
\begin{array}{cc}
h\partial + \partial h & v \partial + \partial v \\
\\
\\
v \partial + \partial v & - s (v \partial + \partial v) \\
\end{array} \right)
\left(
\begin{array}{c}
\delta H^*/\delta h \\
\\ 
\\
\delta H^*/\delta v \\ 
\end{array} \right) ; \tag{4.20$a$}
\end{equation}

(ii) \ The matrix

\begin{equation}
b =
\left(
\begin{array}{cc}
h \partial + \partial h & v \partial + \partial v \\
\\
\\
v \partial + \partial v & - s (v \partial + \partial v) \tag{4.20$b$} 
\end{array} \right)
\end{equation}
is Hamiltonian; 

\noindent
(iii) \ The reduction map $A^*_n = \int\limits^{h}\limits_{0} (v + sy)^n dy $ (2.2) is Hamiltonian between the 
Hamiltonian matrices 
$b$ (4.20$b$) and $B^{2,1}$ (4.6).

\noindent
{\bf{Proof}}. \ (i) \ We have to verify that 

\begin{subequations} \label{eqA}
\begin{gather}
(h \partial + \partial h) (H^*_h) + (v \partial + \partial v) (H^*_v) \ \shortstack{?\\=} \ (A^*_m \partial + 
(2m+1) \partial A^*_m) (H^*_m) ,  \tag{4.21$a$} \\
(v \partial + \partial v) (H^*_h) - s (v \partial + \partial v) (H^*_v) \ \shortstack{?\\=} \ ((2m+1) v^m v_x 
+ 2v^{m+1} \partial)(H^*_m).  \tag{4.21$b$}
\end{gather}
\end{subequations}
Formulae (2.5, 6$b$, 13,) and (3.10) transform the identities (4.21) into the form:

\begin{subequations} 
\begin{gather}
(h \partial + \partial h) \sigma^m + (v \partial + \partial v) \frac{\sigma^m - v^m} {s} \ \shortstack{?\\=} \ 
A^*_m \partial + (2m+1) \partial A^*_m,  \tag{4.22$a$} \\
(v \partial + \partial v) v^m \ \shortstack{?\\=} \ (2m+1) v^m v_x + 2 v^{m+1} \partial. \tag{4.22$b$} 
\end{gather}
\end{subequations}

Formula (4.22$b$) is obvious.

Formula (4.22$a$) can be rewritten as

\begin{equation}
(\sigma \partial + \partial \sigma) \sigma^m - (v \partial + \partial v) v^m \ \shortstack{?\\=} \ \frac
{\sigma^{m+1} - v^{m+1}} {m+1} (2m-2) \partial + (2 m+1) \bigg( \frac {\sigma^{m+1} - v^{m+1}} {m+1} \bigg)_x,  
\tag{4.23}
\end{equation}
and it follows from formula (4.22$b$);

\noindent
\noindent
(ii) \ Let ${\cal{G}}$ be a Lie algebra.  It acts by derivations on itself.  Hence, we can form the semidirect sum 
Lie algebra, with the commutator

\begin{equation}
\left(
\begin{array}{c}
x_1 \\
\\
\\
y_1 
\end{array} \right),
\left(
\begin{array}{c}
x_2 \\
\\
\\
y_2 
\end{array} \right) =
\left(
\begin{array}{c}
[x_1, x_2] \\
\\
\\
\relax
[x_1, y_2] - [x_2, y_1] + \epsilon [y_1, y_2] 
\end{array} \right)  \tag{4.24}
\end{equation}

\noindent
where $\epsilon$ is an arbitrary constant that, when nonzero, can be scaled away to $\epsilon =1$.  The 
Hamiltonian matrix corresponding to the commutator (4.24) has the form

\begin{equation}
\left(
\begin{array}{cc}
B_h & B_v \\
\\
\\
B_v & \epsilon B_v 
\end{array} \right), \tag{4.25}
\end{equation}

\noindent
where $\epsilon = -s$ and $B = B_q = B_q ({\cal{G}})$ is the Hamiltonian matrix naturally attached to the Lie 
algebra ${\cal{G}}$ with the dual coordinates on ${\cal{G}}^*$ denoted by $q$.  Our matrix $b$ (4.20$b$) is of the 
form (4.25), with ${\cal{G}} = {\cal{D}} ({\bf{R}}^1)$ being the Lie algebra of vector fields on ${\bf{R}}^1$; 
\noindent
(iii) \ We have to verify the identity
\begin{equation}
JbJ^t \  \shortstack{?\\=} \ B^{2, 1*} , \tag{4.26}
\end{equation}

\noindent
where the matrix $b$ is given by formula (4.20$b$).  In components, this identity becomes:

\begin{gather}
(A^*_{n,h} (h \partial + \partial h) + A^*_{n,v} (v \partial + \partial v) ) A^*_{m,h} + v^n (v \partial + \partial v) 
A^*_{m,v} \  \shortstack{?\\=} \notag \\
 \shortstack{?\\=} \ (2n+1) A^*_{n+m} \partial + \partial (2m+1) A^*_{n+m}. \tag{4.27}
\end{gather}

By formulae (2.2-4), this can be rewritten as
\begin{gather}
(\sigma^n (h \partial + \partial h) + \frac {\sigma^n - v^n} {s} (v \partial + \partial v) \sigma^m + v^n (v \partial 
+ \partial v) \frac {\sigma^m - v^m} {s} \  \shortstack{?\\=} \notag \\
 \shortstack{?\\=}  \ (2n+1) \frac {\sigma^{n+m+1} - v^{n+m+1}} {(n+m+1)s} \partial + \partial (2m+1) 
 \frac {\sigma^{n+m+1} - v^{n+m+1}} {(n+m+1)s} = \notag \\
= \frac {2} {s} (\sigma^{n+m+1} - v^{n+m+1} ) \partial + \frac{2m+1} {s} (\sigma^{n+m} \sigma_x - v^{n+m} v_x), 
\tag{4.28}
\end{gather}
or as
\begin{gather}
(\sigma^n (\sigma \partial + \partial \sigma) - v^n (v \partial + \partial v) ) \sigma^m + v^n (v \partial + 
\partial v) (\sigma^m - v^m) = \notag \\
= \sigma^n (\sigma \partial + \partial \sigma) \sigma^m - v^n (v \partial + \partial v) v^m \ \shortstack{?\\=} \notag 
\\
\shortstack{?\\=} 2 (\sigma^{n+m+1} - v^{n+m+1} ) \partial + (2m+1) (\sigma^{n+m} \sigma_x - v^{n+m} 
v_x, \tag{4.29}
\end{gather}

\noindent
which follows from the single equality

\begin{equation}
w^n (w \partial + \partial w) w^m = 2 w^{n+m+1} \partial + (2m+1) w^{n+m} w_x, \tag{4.30}
\end{equation}

which in turn follows from formula (4.22$b$) or is obvious in its own right $\blacksquare$

 \noindent
{\bf{Remark 4.31}}. \ Formulae in [4,5] suggest - but not prove - that the constant-shear reductions of Hamiltonian 
systems do not exist in the $N+1$ dimension for $N \not= 1$.

\section{Lie Algebra Of Reducible Flows}

A reducible dynamical system (4.2) in the momentum space has, by formula (4.5), the form 
\begin{equation}
\hat X (A_n) = A_{n,t} = (n + m) A_{n+m-1} \bar P_m + A_{n+m,x} P_m. \tag{5.1}
\end{equation}
Suppose we have another reducible vector field, $\hat Y$:
\begin{equation}
\hat Y (A_n) = (n+m) A_{n+m-1} \bar \Phi_m + A_{n+m,x} \Phi_m. \tag{5.2}
\end{equation}
{\bf{Theorem 5.3}}.  The commutator of reducible vector fields is again reducible.
\noindent
\\
{\bf{Proof}}.  We shall show that
\begin{equation}
[\hat X, \hat Y] (A_n) = (n+r) A_{n+r-1} \bar \Omega_r + A_{n+r, x} \Omega_r, \tag{5.4}
\end{equation}
where
\begin{gather}
\bar \Omega_r = \hat X (\bar \Phi_r) - \hat Y (\bar P_r) + \sum_{k+m=r+1} \bar \Phi_k \bar P_m + \notag \\
+ \sum_{k+m=r} (\Phi_k \bar P_{m,x} - P_k \bar \Phi_{m,x}), \tag{5.5$a$} \\
\Omega_r = \hat X (\Phi_r) - \hat Y (P_r) + \sum_{k+m=r+1} (k \Phi_k \bar P_m - m P_m \bar \Phi_k) + \notag \\
+ \sum_{k+m=r} (\Phi_k P_{m,x} - P_m \Phi_{k,x}). \tag{5.5$b$}
\end{gather}
We have:
\begin{gather}
[\hat X, \hat Y] (A_n) - (n+r) A_{n+r-1} [\hat X (\bar \Phi_r) - \hat Y (\bar P_r)] - \tag{5.6$a$} \\
- A_{n+r,x} [\hat X (\Phi_r) - \hat Y (P_r)] = \tag{5.6$b$} \\
= \bar \Phi_k (n+k) \hat X (A_{n+k-1}) + \Phi_k [\hat X (A_{n+k})]_x - \notag \\
- \bar P_m (n+m) \hat Y (A_{n+m-1}) - P_m [\hat Y (A_{n+m})] _x = \notag\
\end{gather}

\begin{gather}
= \bar \Phi_k (n+k) [(n+k-1+m) A_{n+k-1+m-1}) \bar P_m + A_{n+k-1+m,x} P_m] + \tag{5.7$a$} \\
+ \Phi_k [(n+k+m) (A_{n+k+m-1,x} \bar P_m + A_{n+k+m-1} \bar P_{m,x}) + \tag{5.7$b$} \\
+ A_{n+k+m, xx} P_m + A_{n+k+m,x} P_{m,x}] - \tag{5.7$c$} \
\end{gather}

\begin{gather}
- \bar P_m (n+m) [(n+m-1+k) A_{n+m-1+k-1} \bar \Phi_k + A_{n+m-1+k, x} \bar \Phi_k ] - \tag{5.8$a$} \\
- P_m [(n+m+k) (A_{n+m+k-1, x} \bar \Phi_k + A_{n+m+k-1} \bar \Phi_{k,x}) + \tag{5.8$b$} \\
+ A_{n+m+k,xx} \Phi_k + A_{n+m+k, x} \Phi_{k,x}]. \tag{5.8$c$} \
\end{gather}

The first summands in (5.7$c$) and (5.8$c$) cancel out.

The second summands in (5.7$a$) and (5.8$a$), and the first summands in (5.7$b$) and (5.8$b$), combine into 

\begin{equation}
A_{n+r,x} \sum_{k+m=r+1} (k \Phi_k \bar P_m - m P_m \bar \Phi_k), \tag{5.9}
\end{equation}
while the second summands in (5.7$c$) and (5.8$c$) yield
\begin{equation}
A_{n+r,x} \sum_{k+m=r} (\Phi_k P_{m, x} - P_m \Phi_{k,x}) . \tag{5.10}
\end{equation}
Formulae (5.6$b$, 9, 10) account for formula (5.5$b$).

What remains, the first summands in (5.7$a$) and (5.8$a$), and the second summands in (5.7$b$) and (5.8$b$), combine 
into
\begin{gather}
(n+r) A_{n+r-1} \{ \sum_{k+m=r+1} [(n+k) \bar \Phi_k \bar P_m - (n+m) \bar P_m \bar \Phi_k] + \notag \\
+ \sum_{k+m=r} (\Phi_k \bar P_{m,x} - P_m \bar \Phi_{k,x}) \} \tag{5.11} \
\end{gather}
which, together with the second summand in (5.6$a$), account for formula (5.5$a$) $\blacksquare$

Since the momentum map
\begin{equation}
A_n = \int^h_0 u^n dy, \ \ \ n \in {\bf{Z}}_{\geq 0}, \tag{5.12}
\end{equation}
is injective, Theorem 5.3 implies that the 2+1-dimensional hydrodynamic systems (4.1) that are reducible in the 
physical space, 
\begin{gather}
h_t = (m+1) \bar P_{m+1} A_m + P_m A_{m,x} , \tag{5.13$a$} \\
u_t = P_m u^m u_x + \bar P_m u^m - u^y \int^y_0 dy (P_m (u^m)_x + (m+1) \bar P_{m+1} u^m), \tag{5.13$b$} \
\end{gather}
also form a Lie algebra.  In particular, when there is no $x$-dependence, so that the $P_m$'s vanish and the $\bar 
P_m$'s 
depend on the $A_n$'s but not on the derivatives of the $A_n$'s, we get a free-surface analog of the Lie algebra 
ODE's:
\begin{gather}
h_t = (m+1) \bar P_{m+1} A_m, \tag{5.14$a$} \\
u_t = \bar P_0 +  \bar P_{m+1} (u^{m+1} - u_y \int^y_0 dy (m+1) u^m). \tag{5.14$b$} \
\end{gather}
All these Lie algebras possess subalgebras induced by the constain
\begin{equation}
\{h=A_0 = A_1 = ... = 0 \}. \tag{5.15}
\end{equation}
{\bf{Remark 5.16}}.  The 2+1-dimensional free-surface hydrodynamics is infused with Lie algebras, such as (4.1), 
(4.2), (5.1), (5.13).  This picture is very different from the theory of systems of hydrodynamical type in $1+1-d$, of 
the form
\begin{equation}
u_{i,t} = \sum_j {\cal{M}}_i^j (\mbox{\boldmath $u$})u_{j,x} , \tag{5.17}
\end{equation}
where, in general, a commutator of two such systems is no longer of hydrodynamic type (see [8]).

\section{Multilayer Representations}

Imagine that the Benney system (1.5) 

\begin{gather}
h_t = (\int\limits^{h}\limits_{0} udy)_x, \tag{6.1$a$} \\
u_t = uu_x + gh_x - u_y \int\limits^{y}\limits_{0} u_x dy, \tag{6.1$b$} \  
\end{gather}
is broken into $N$ layers
\begin{equation}
h_{k-1} \leq y \leq h_k, \ \ \ k =1, ..., N, \ h_0 = 0, \tag{6.2}
\end{equation}
such that in each layer the velocity profile $u_k$ is $y$-independent.  The Benney system them turns into the 
2$N$-component system
\begin{gather}
h_{k,t} = (u_k h_k)_x, \tag{6.3$a$} \\
u_{k,t} = u_k u_{k,x} + gh_x, \ \ \ k = 1, ..., N, \tag{6.3$b$} \\
h = \sum^{n}_{k=1} h_k. \tag{6.3$c$} 
\end{gather}
This idea and the system (6.3) are due to Zakharov [10], who in addition showed that, rather mysteriously, the system 
(6.3) appears also as the zero-dispersion limit of a vector Nonlinear Schr${\ddot{\rm{o}}}$dinger equation.

The system (6.3) can be considered as a multi-component version of the classical long-wave system (1.1), and it was 
analyzed in detail by Pavlov and Tsarev [9].  As Zakharov noted, the moment map (1.6)

\begin{equation}
A_n =  = \int\limits^{h}\limits_{0} u^n  dy, \ \ \ n \in {\bf{Z}}_{\geq 0}, \tag{6.4}
\end{equation}
which now becomes 
\begin{equation}
A_n = \sum^N_{k=1} h_k u_k^n, \ \ \ n \in {\bf{Z}}_{\geq 0}, \tag{6.5}
\end{equation}
maps the 2$N$-component system (6.3) into the infinite-component Benney momentum system (1.7)

\begin{equation}
A_{n,t} = A_{n+1,x} + gn A_{n-1} A_{0,x}, \ \ \ n \in {\bf{Z}}_{\geq 0}. \tag{6.6}
\end{equation}
 
 It's easy to see that the same conclusion applies to all the higher Benney flows constructed in [6, 7], and indeed 
 to {\it{any}} flow (1.9) in the Hamiltonian structures (1.10), (4.7) or any other {\it{linear}} Hamiltonian 
structure.  The argument is as follows.

Let ${\cal{G}}$ be a Lie algebra, $C_A$ the (differential-difference) ring of functions on ${\cal{G}}^*$, and $B = 
B_A$ the natural Hamiltonian structure in $C_A$ attached to ${\cal{G}}$ (see [5].)  Let ${\cal{G}}^{<N>}$ be the 
direct sum of $N$ copies of ${\cal{G}}$.  The homomorphism of Lie algebras
\begin{gather}
\varphi: {\cal{G}}^{<N>} \ \rightarrow \ {\cal{G}}, \tag{6.7$a$}\\
\varphi \bigg(\oplus^N_{k=1} x_k\bigg) = \sum^N_{k=1} x_k, \tag{6.7$b$} \
\end{gather}
induces the corresponding linear Hamiltonian map

\begin{gather}
\Phi: C_A \ \rightarrow \ C_A^{<N>}, \notag \\
\Phi (A_n) = \sum^N_{k=1} A_{n|k}. \tag{6.8} 
\end{gather}
If $C_V$ is another ring, with a Hamiltonian structure on it, and if
\begin{equation}
w: C_A \ \rightarrow \ C_V \tag{6.9}
\end{equation}
is a Hamiltonian map, then so is its $k^{th}$-copy version:
\begin{equation}
w_k: C_{A_{|k}} \ \rightarrow \ C_{V_{|k}}. \tag{6.10}
\end{equation}
The composition
\begin{gather}
Z = \Phi \circ \Omega: \ C_A \ \rightarrow \ C_V^{<N>}, \tag{6.11} \\
\Omega = \oplus^N_{k=1} w_k, \ C_V^{<N>} = \oplus^N_{k=1} C_{V_{|k}}, \tag{6.12} \
\end{gather}
is then a multilayer analog of the single-layer canonical map (6.9).

In particular, for the Hamiltonian matrix (1.10)
\begin{gather}
h_t = \partial (m A_{m-1} H_m) \tag{6.13$a$}\\
u_t = \partial (u^{m} H_m) - u_y \int\limits^{y}\limits_{0} dy (mu^{m-1} H_m)_x. \tag{6.13$b$}\  
\end{gather}
with the Hamiltonian map (1.13)
\begin{equation}
w (A_n) = hu^n, \tag{6.14}
\end{equation}
the general construction above gives:
\begin{gather}
h_{k,t} = \partial \frac{\delta}{\delta u_k} Z (H), \tag{6.15$a$} \\
u_{k,t} = \partial \frac{\delta}{\delta h_k} Z (H), \tag{6.15$b$} \\
Z(A_n) = \sum^N_{k=1} h_k u_k^n. \tag{6.15$c$}\
\end{gather}
If $H$ is the 2$^{nd}$ Hamiltonian (1.8), 
\begin{equation}
H = \frac{1}{2} (A_2 + g A^2_0) \tag{6.16}
\end{equation}
then
\begin{gather}
Z (H) = \frac{1}{2} (\sum_k h_k u^2_k + g (\sum_k h_k)^2), \tag{6.17$a$} \\
\frac{\delta Z (H)}{\delta u_k} = h_k u_k, \ \ \ \frac{\delta Z(H)}{\delta h_k}  = \frac{u_k^2}{2} + gh, \tag{6.17$b$} 
\
\end{gather}
and we recover the Zakharov system (6.3).  If $H$ is the $3^{rd}$ Hamiltonian,
\begin{equation}
H = \frac{1}{3} A_3 + g A_0 A_1, \tag{6.18}
\end{equation} 
then
\begin{gather}
Z (H) = \frac{1}{3} \sum_k h_k u_k^3 + g (\sum_k h_k u_k) (\sum_\ell h_\ell) , \tag{6.19$a$} \\
\frac{\delta Z (H)}{\delta u_k} = h_k u_k^2, \ \frac{\delta Z(H)}{\delta h_k} = \frac{u_k^3}{3} + gu_k h + g A_1 
\Rightarrow \tag{6.19$b$} \
\end{gather}

\begin{gather}
h_{k,t} = (h_k u_k^2)_x, \tag{6.20$a$} \\
u_{k, t} = u_k^2 u_{k,x} + g (u_k h + A_1)_x,  \ \ \ k =1, ..., N. \tag{6.20$b$} \
\end{gather}
All the odd-numbered higher Benney flows can be restricted onto the invariant submanifold [3]
\begin{equation}
\{A_1 = A_3 = A_5 = ... = 0\}. \tag{6.21}
\end{equation}
The analog of this fact is this:  Suppose $N$ is even:
\begin{equation}
N = 2M , \tag{6.22$a$}
\end{equation}
and
\begin{equation}
h_{k+m} = h_k, \ u_{k+m} = - u_k,  \ \ k = 1, ..., M. \tag{6.22$b$}
\end{equation}
Then all the odd flows in the $(h, u)$-space can be properly reduced by the constrain (6.22).  The 3$^{rd}$ flow 
(6.20) becomes:
\begin{gather}
h_{k,t} = (h_k u_k^2)_x, \tag{6.23$a$}\\
u_{k,t}= u_k^2 u_x + 2g (u_x h)_x, \ \ \ k=1,..., M , \tag{6.23$b$}
\end{gather}
in the variables
\begin{equation}
h_k = h_k, \ U_k = u^2_k, \ \ \ k =1, ..., M, \tag{6.24}
\end{equation}

\begin{gather}
h_{k,t} = (h_k U_k)_x, \tag{6.25$a$} \\
U_{k,t} = U_k U_{k,x} + 2g (U_{k,x} h + 2 U_k h_x), \ \ \ k =1, ..., M. \tag{6.25$b$} \
\end{gather}

Similar Hamiltonian construction applies to the case of constant shear.  Equations (6.15) become
\begin{gather}
h_{k,t} = \partial \frac{\delta}{\delta v_k} Z(H) , \tag{6.26$a$} \\
v_{k,t} = \partial \bigg( \frac{\delta}{\delta h_k}  - s_k \frac{\delta}{\delta v_k} \bigg) Z(H), \ \ \ k =1, ..., N, 
\tag{6.26$b$} \
\end{gather}
where now
\begin{equation}
Z(A_n) = \sum_k \int^{h_k}_0 (v_k + s_k y)^n dy = \sum_k \frac{(v_k + s_k h_k)^{n+1} - v_k^{n+1}}{s_k (n+1)} . 
\tag{6.27}
\end{equation}
In particular, for the 2$^{nd}$ flow, we get:
\begin{gather}
h_{k,t} = (h_k v_k + \frac{s_k}{2} h^2_k)_x, \tag{6.28$a$} \\
v_{k,t} = v_k v_{k,x} + gh_x, \ \ \ k = 1, ..., N. \tag{6.28$b$}\
\end{gather}
This is a multicomponent version of the single component shear system (1.18), and a shear extension of the Zakharov 
system (6.3).

The shear analog of the constrain (6.22) now becomes:
\begin{equation}
h_{k+m} = h_k, \ \ v_{k+m} = - v_k, \ \ s_{k+M} = - s_k, \ \ \ k=1, ..., M. \tag{6.29}
\end{equation}

The construction above works only for systems with {\it{linear}} Hamiltonian structures.  It doesn't apply to the 
hydrodynamic chain [3]
\begin{gather}
A_{n,t} = A_{n+1,x} + (an +b) A_n A_{0,x} + c A_0 A_{n,x} , \ n \in {\bf{Z}}_{\geq 0}, \tag{6.30$a$} \\
a, b, c = const, \tag{6.30$b$} \
\end{gather}
with
\begin{equation}
b \neq 2c. \tag{6.30$c$}
\end{equation}
Nevertheless, we now show that the Zakharov map (6.5) and its shear version (6.27) work in the most general 
circumstances.  
\\
{\bf{Theorem 6.31}}. (i) The Zakharov map (6.5) applies to {\it{any}} flow (4.2):

\begin{equation}
A_{n,t} = n A_{n+m-1} \bar P_m + A_{n+m} Q_m + A_{n+m,x} P_m. \tag{6.32}
\end{equation}
(ii) The map (6.27) applies to {\it{any}} flow (5.1):
\begin{equation}
A_{n,t} = (n + m) A_{n+m-1} \bar P_m + A_{n+m,x} P_m. \tag{6.33}
\end{equation}
{\bf{Proof}}.  The idea is this.  Either of the two maps is a local diffeomorphism between the variables $h_1, u_1, 
..., h_N, u_N$ and $A_0, ..., A_{2N-1}$.  Therefore, we can consider these maps as providing an anvariant submanifold 
for each flow.  This means that if we find $-$ by whatever means $-$ an evolution in the $\{u_k, h_k\}-$space that is 
properly embedded into the evolution in the $A-$space, that's it.
\\

Now for the details.  Writing $(\cdot)^{\cdot}$ instead of $(\cdot)_t$, we convert, first for the Zakharov case, 
formula (4.2) into
\begin{gather} 
\sum_k (\dot h_k u_k^n + nh_k u_k^{n-1} \dot u_k) = \notag \\
= \sum_k \{n h_k u_k^{n+m-1} \bar P_m^* + h_k u_k^{n+m} Q^*_m + [h_{k,x} u_k^{n+m} + (n+m)h_k u_k^{n+m-1} u_{k,x}] 
P_m^* \}, \tag{6.34}\ 
\end{gather}
where
\begin{equation}
( \cdot )^* = Z( \cdot ), \ Z(A_n) = \sum_k h_k u_k^n, \ \ n \in {\bf{Z}}_{\geq 0}. \tag{6.35} 
\end{equation}
We drop the $\sum_k$ operator from each side, multiply the resulting equation by $u_k^{-n}$, and find:
\begin{gather}
\dot h_k + nh_k u_k^{-1} \dot u_k = \notag \\
= h_k u_k^m Q_m^* + [h_{k,x} u_k^m + m h_k u_k^{m-1} u_{k,x}] P_m^* + \tag{6.36$a$} \\
+n h_k u_k^{-1} [u_k^m \bar P_m^* + u_k^m u_{k,x} P_m^*], \tag{6.36$b$} \
\relax
\end{gather}
whence
\begin{gather}
\dot h_k = h_k u_k^m Q_m^* + [h_{k,x} u_k^m + m h_k u_k^{m-1} u_{k,x}] P_m^*, \tag{6.37$a$} \\
\dot u_k = u_k^m \bar P_m^* + u_k^m u_{k,x} P_m^*. \tag{6.37$b$} \
\relax
\end{gather}

Next, denoting
\begin{equation}
\sigma_k = v_k + s_k h_k, \tag{6.38}
\end{equation}
so that
\begin{equation}
Z (A_n) = \sum_k \frac{\sigma_k^{n+1}- v_k^{n+1}}{(n+1)s_k} , \tag{6.39}
\end{equation}
we convert formula (5.1) into
\begin{gather}
{\dot {A}_n} = \sum_k \frac{(\sigma_k^{n+1} - v_k^{n+1})^\cdot}{(n+1)s_k} = \sum_k \{ \sigma_k^n \dot h_k + 
\frac{\sigma_k^n - v_k^n}{s_k} \dot v_k \} = \notag \\
= \sum_k \{ \frac{\sigma_k^{n+m} - v_k^{n+m}}{s_k} \bar P_m^* + \bigg(\frac{\sigma_k^{n+m+1} - v_k^{n+m+1}} 
{(n+m+1)s_k} \bigg)_x P_m^*\}, \tag{6.40} \
\relax
\end{gather}
where again
\begin{equation}
(\cdot)^* = Z(\cdot), \tag{6.41}
\relax
\end{equation}
but with the map $Z$ given by formula (6.39).

We now drop the operator $\sum_k$ from each side of formula (6.40) and get, for each $k$, the equation 
\begin{gather}
 \sigma_k^n \dot h_k + \frac{\sigma_k^n - v_k^n}{s_k} \dot v_k  = \notag \\
=   \frac{\sigma_k^{n+m} - v_k^{n+m}}{s_k} \bar P_m^* + \bigg(\frac{\sigma_k^{n+m+1} - v_k^{n+m+1}} {(n+m+1)s_k} 
\bigg)_x P_m^*, \tag{6.42} \
\end{gather}
For $n=0$, equation (6.42) yields:
\begin{equation}
\dot h_k = \frac{\sigma_k^m - v_k^m}{s_k} \bar P_m^* + \bigg(\frac{\sigma_k^{m+1} - v_k^{m+1}}{(m+1)s_k}\bigg)_x 
P_m^*. \tag{6.43}
\relax
\end{equation}
Substituting this back into equation (6.42), we find:
\begin{equation}
\frac{\sigma_k^n - v_k^{n}}{s_k} \dot v_k = [\sigma_k^{n+m} - v_k^{n+m} - \sigma_k^n (\sigma^m_k - v_k^m)] s_k^{-1} 
\bar P_m^* + \notag \
\end{equation}

\begin{equation}
+ [ (\sigma_k^{n+m} - \sigma_k^n \sigma_k^m) \sigma_{k,x} - (v_k^{n+m} - \sigma_k^n v_k^m)v_{k,x}] s_k^{-1} P_m^* 
\Leftrightarrow \tag{6.44}
\end{equation}

\begin{equation}
\dot v_k = v_k^m \bar P_m^* + v^m_k v_{k,x} P^*_m \ \ \blacksquare \tag{6.45}
\end{equation}


\begin{thebibliography}{10}

\bibitem{}
	Benney, D. J. {\it{Some Properties of Long Nonlinear Waves}}, Stud. Appl. Math. {\bf{L11}} (1973) 45-50.
	
\bibitem{}
	Kupershmidt, B. A., {\it{Deformations of Integrable Systems}}, Proc. Roy. Irish Acad. {\bf{83}} A. No. 1 (1983) 
	45-74.

\bibitem{}
	Kupershmidt, B. A., {\it{Normal and Universal Forms in Integrable Hydrodynamical Systems in Proc. of NASA 
Ames-Berkley Conf. on Nonlinear Problems in Optimal Control and Hydrodynamics}}, L. R. Hunt and C. F. Martin Ed-s, 
Math. Sci. Press (1984) 357-378.	
	
\bibitem{}
	Kupershmidt, B. A., {\it{Hydrodynamical Poisson Brackets and Local Lie Algebras}}, Phys. Lett. {\bf{121A}} 
	(1987) 167-174.

\bibitem{}
	Kupershmidt, B. A., {\it{The Variational Principles of Dynamics}}, World Scientific (Singapore, 1992).
	
\bibitem{}
	Kupershmidt, B. A. and Manin, Yu I.,  {\it{Long-Wave Equation with Free Boundaries. I. Conservation Laws}}, Funct. 
	Anal. App. {\bf{11}}:3 (1977) 31 - 42 (Russian); 188-197 (English).	

\bibitem{}
 	Kupershmidt, B. A. and Manin, Yu I., {\it{Equations of Long Waves with a Free Surface.  II Hamiltonian Structure 
and Higher Equations}}, Funct. Anal. Appl. {\bf{12}}:1 (1978) 25-37 (Russian); 20-29 (English).

\bibitem{}
	Pavlov, M. V., Svinolupov, S. I., and Sharipov, R. A., {\it{ An Invariant Criterion for Hydrodynamic 
Integrability}}, Funktsional. Anal. i Prilozhen.  {\bf{30}} no. 1 (1996) 18--29, 96 (Russian);  
Funct. Anal. Appl. {\bf{30}} no. 1 (1996) 15--22 (English);
arXiv:solv-int/9407003. 

\bibitem{}
	Pavlov, M. V., and Tsarev, S. P., {\it{Conservation Laws for the Benney Equations}},   Uspekhi Mat. Nauk  
{\bf{46}}  no. 4 (1991) 169--170 (Russian);   Russian Math. Surveys  {\bf{46}} no. 4 (1991) 196--197 (English).

\bibitem{}
	 Zakharov, V. E., {\it{Benney Equations and Quasiclassical Approximation in the Inverse Problem Method}}, 
Funktsional. Anal. i Prilozhen. {\bf{14}} no. 2 (1980) 15--24 (Russian); Functional Anal. Appl. {\bf{14}} no. 2 (1980)
89--98 (English). 



\end{thebibliography}
\end{document}